\documentclass[prd, aps, superscriptaddress, preprintnumbers, floatfix, nofootinbib]{revtex4}

\usepackage{amsfonts}
\usepackage{amsmath}
\usepackage{amssymb}
\usepackage{bm}
\usepackage{dcolumn}
\usepackage{epsfig}
\usepackage{graphicx}   
\usepackage{graphics}
\usepackage[latin1]{inputenc}
\usepackage{latexsym}
\usepackage{rotating}
\usepackage{hyperref}
\usepackage{color}

\newcommand\be{\begin{equation}}
\newcommand\ba{\begin{eqnarray}}
\newcommand\ee{\end{equation}}
\newcommand\ea{\end{eqnarray}}

\begin{document}

\title{Dynamics of Cosmological Perturbations and Reheating in the Anamorphic Universe}

\author{L. L. Graef}
\email{leilagraef@on.br}
\affiliation{Physics Department, McGill University, Montreal, QC, H3A 2T8, Canada}
\affiliation{Observat\'orio Nacional, 20921-400, Rio de Janeiro - RJ, Brazil}

\author{W. S. Hip\'olito-Ricaldi}
\email{wiliam.ricaldi@ufes.br}
\affiliation{Departamento de Ci\^encias Naturais, Universidade Federal do Esp\'{\i}rito Santo, Rodovia BR 101 Norte, km. 60, S\~ao Mateus, ES, Brazil}

\author{Elisa G.M. Ferreira}
\email{elisa.ferreira@mail.mcgill.ca}
\affiliation{Physics Department, McGill University, Montreal, QC, H3A 2T8, Canada}

\author{Robert Brandenberger}
\email{rhb@physics.mcgill.ca}
\affiliation{Physics Department, McGill University, Montreal, QC, H3A 2T8, Canada}

\date{\today}

%%%%%%%%%%%%%%%%%%%%%%%%%%%%%%%%%%%%%%%%%%%%%%%%%%%%%%%%%%%%%%%%%%%%%%%%%%%%%%%%%%%%%%%%%%%%%%
\begin{abstract}
We discuss scalar-tensor realizations of the Anamorphic cosmological scenario recently proposed by  
Ijjas and Steinhardt \cite{an}. Through an analysis of the dynamics of cosmological perturbations we obtain
constraints on the parameters of the model. We also study gravitational Parker particle production in the
contracting Anamorphic phase and we compute the fraction between the energy density
of created particles at the end of the phase and the background energy
density. We find that, as in the case of inflation, a new mechanism is required to reheat
the universe.
\end{abstract}
%%%%%%%%%%%%%%%%%%%%%%%%%%%%%%%%%%%%%%%%%%%%%%%%%%%%%%%%%%%%%%%%%%%%%%%%%%%%%%%%%%%%%%%%%%%%%%

\maketitle

%%%%%%%%%%%%%%%%%%%%%%%%%%%%%%%%%%%%%%%%%%%%%%%%%%%%%%%%%%%%%%%%%%%%%%%%%%%%%%%%%%%%%%%%%%%%%%
\section{Introduction} 

In the last years we have witnessed the discovery of a lot of precision data concerning the
structure of the universe on large scales (see e.g. \cite{planck, planckbissep} for recent
cosmic microwave background (CMB) anisotropy results). This data can only be explained
by invoking a mechanism of structure formation operating in the early universe
(see e.g. \cite{SZ} and \cite{PY}). On the other hand, data on the large-scale structure
of the universe will then also allow us to probe the physics of the very early universe.

Inflation is the current paradigm for early universe cosmology. The inflationary scenario attempts 
to solve problems of Standard Big Bang cosmology such as the {\it horizon} and {\it flatness} problems
by invoking a period of rapid accelerated expansion of space \cite{Guth} (see 
also \cite{Brout, Sato, Fang, Kazanas, Starob1}). Inflation provided a causal explanation for the
origin of CMB anisotropies and the large-scale structure of the universe \cite{Mukh} (see
also \cite{Press}). The inflationary scenario was, in fact, developed before most of the
data we now have was in hand. Hence, inflation has been a very predictive scenario. 
However, inflation also suffers from several problems (see e.g. \cite{RHBrev} for
an early review). For example, if initial conditions are set at the time when the
energy density was given by the Planck scale, then the initial spatial curvature must be tuned 
to be small in order to enter an inflationary phase (the energy density at that point being
several orders of magnitude smaller). In the inflationary scenario there is also a
{\it trans-Planckian problem} for fluctuations - unless the inflationary phase only lasts
close to the minimal amount of time it must in order for the scenario to solve the horizon
problem, then all scales which we currently observed had a wavelength smaller than
the Planck length at the beginning of inflation, and hence the applicability of Einstein
gravity and standard matter actions to the initial development of fluctuations is
questionable \cite{Jerome}.

Inflation is not the only early universe scenario compatible with cosmological observations 
\cite{RHBrevs, peternelson1, nelson}. There are several alternatives which solve the problems of 
Standard Big Bang cosmology and which also produce the observed primordial spectrum of fluctuations. 
One example are bouncing models that involve an initial matter-dominated phase of contraction 
\cite{MBrev}. In such models, there is no trans-Planckian problem for fluctuations as long
as the energy scale of the bounce is smaller than the Planck scale. However, as emphasized 
in \cite{Peter2},  these scenarios suffer from an instability in the
contracting phase to the growth of anisotropies.  Among the alternatives to inflation, one which 
is able to avoid the anisotropy problem, solve the trans-Planckian problem
for fluctuations and in which the homogeneous and isotropic background trajectory
is an attractor in initial condition space is the {\it Ekpyrotic scenario} \cite{Ekp}. Ekpyrotic cosmology 
assumes an early period of ultra-slow contraction which renders 
the universe spatially flat, smooth (modulo quantum vacuum fluctuations) and isotropic \cite{NoBKL},
followed by a cosmological bounce which leads to a transition to the expanding phase of
Standard Big Bang cosmology. Like in inflationary cosmology, it is assumed that fluctuations
originate as quantum vacuum perturbations on sub-Hubble scales which get squeezed
and decohere as scales exit the Hubble radius during the contracting phase. However,
the induced curvature perturbations obtain a spectrum which is nearly vacuum and hence
far from scale-invariant \cite{Lyth,FB} unless entropy fluctuations are invoked \cite{EkpEnt}.
Even if entropy fluctuations are introduced, the predicted spectrum of gravitational radiation
is highly blue, and generically large non-Gaussianities are induced (see e.g. \cite{EkpNG}
for a review of this topic).

Recently, a new model for the early universe called {\it Anamorphic cosmology} was proposed by Ijjas 
and Steinhardt  \cite{an,an2,anna}. This scenario combines elements and advantages of both 
the inflationary and the Ekpyrotic scenarios. It is based on the realization that in scalar-tensor
theories of gravity such as dilaton gravity (the low energy limit of string theory) the gravitational
parameter is time-dependent in a frame in which the parameters of the matter action (e.g. the
mass $m$ of a particle) are constant (the {\it string frame}), and on the other hand the matter parameters vary in
the frame in which the gravitational parameter (or equivalently the Planck mass $M_{Pl}$) 
is constant (the {\it Einstein frame}) \footnote{Note that this point was
the crucial one in the {\it Pre-Big-Bang cosmology} scenario developed in the early
1990s \cite{PBB}.}. The two frames are related via a Weyl transformation.
The assumption made in the Anamorphic scenario is that the initial
phase of the universe corresponds to Ekpyrotic contraction in the  string frame
(the frame in which the mass of matter particles is constant), 
and to inflationary expansion in the Einstein frame. It
describes a phase (the {\it Anamorphic phase})
in which the universe has a smoothing contracting behavior in the
string frame, whereas cosmological fluctuations and gravitational waves evolve as in
inflationary cosmology. Because the impression of the cosmological background depends on
the frame, the authors referred to this class of models as {\it Anamorphic}. This requires
$m/M_{Pl}$ to decrease at a particular rate. Then both the spatial curvature
and the anisotropies are suppressed, and therefore chaotic mixmaster behavior is avoided, as it
is in Ekpyrotic cosmology \cite{NoBKL}. On the other hand, like in inflationary cosmology
and unlike in the Ekpyrotic scenario, the Anamorphic scenario can directly generate a nearly 
scale-invariant spectrum of adiabatic cosmological perturbations and gravitational waves
using only a single matter scalar field  \footnote{Some of these elements were already considered 
in other works  \cite{Piao, Wetterich, MLi, Dom}.}. Note that after the string frame bounce,
in the Standard Big Bang cosmology phase of expansion, the string and Einstein
frames must coincide. In the following we will 
explicitly compute the power spectrum of cosmological perturbations in a particular class of 
Anamorphic models which will allow us to obtain constraints on the model parameters.

Another important aspect to be analyzed in this new scenario is how the late time thermal state emerges
(the {\it reheating} process).
In inflationary cosmology, it is typically assumed that there is a separate reheating mechanism 
which is a consequence of a coupling between the inflaton field (the scalar field driving inflationary
expansion) and the Standard Model fields. As studied in \cite{DK, TB, STB, KLS2, RehRevs}, the 
energy transfer from the inflaton to Standard Model fields is usually very rapid on Hubble time
scales and proceeds via a parametric instability in the equation of motion for
the Standard Model fields in the presence of an evolving inflaton field.
 On the other hand, it was shown in \cite{Jerome2} that in the {\it matter bounce} scenario, 
 gravitational particle production \cite{Parker} is sufficient to produce
a hot early universe. Recently, it was shown that this same mechanism, under some conditions, can also be responsible 
for the emergence of  a hot thermal state in the {\it New Ekpyrotic Model} \cite{NewEkp}, 
avoiding the need to introduce an additional reheating phase \cite{us}. 
\footnote{Particle creation in bouncing cosmologies was also considered in \cite{nelsonnew} but in 
a different context.}

As previously mentioned, the Anamorphic cosmology combines elements of both inflationary and 
Ekpyrotic scenarios. In view of this 
duality, we are interested in analyzing how the generation of the hot universe proceeds in this model. 
Thus,  following the lines of our recent work \cite{us}, we  compute the energy density produced through 
the Parker mechanism in the Anamorphic scenario. We only study the contribution  to particle
production in the Anamorphic phase of contraction, and neglect additional particle production
which will occur during the bounce. Hence, we will find a lower bound on the total number of particles 
produced during the entire cosmological evolution. 
The goal of this analysis is to understand under which conditions on the model parameters gravitational 
particle production can be sufficient to reheat the universe, eliminating the need to introduce an extra 
reheating mechanism, like is done in inflation. 

This paper is organized as follows. In Section II, we review the Anamorphic scenario and analyze the 
background dynamics  in a specific class of realizations of this scenario. In Section III, we analyze 
the dynamics of the cosmological perturbations in the class of models which we consider. We then study 
Parker particle production during the Anamorphic phase, evaluating the density of produced particles and comparing it to the background density. We conclude in Section V with a discussion. 
We will consider background cosmologies described by a Friedmann-Robertson-Walker metric
with scale factor $a(t)$ and linearized fluctuations about such a metric. 

\section{The Anamorphic Universe}

We begin with a review of the Anamorphic scenario \cite{an}. We consider a theory containing
matter with a characteristic mass $m$ (e.g. the mass of a scalar matter field) and with a 
gravitational constant given by a Planck mass $M_{Pl}$, and we assume that both can be
time-dependent. The condition to obtain a smoothing contracting phase from the point of
view of the string frame (the frame in which $m$ is independent of time), and in which
the Universe is undergoing accelerated expansion from the point of view of the Einstein
frame (the frame in which $M_{Pl}$ is constant) can be expressed in terms of frame-invariant
dimensionless quantities
\begin{eqnarray}
\Theta_m \, = \, \left(H+\frac{\dot{m}}{m}\right)M^{-1}_{Pl}\,, 
\end{eqnarray}
and
\begin{eqnarray}\label{thetaplanck}
\Theta_{Pl} \, = \, \left(H+\frac{\dot{M_{Pl}}}{M_{Pl}}\right)M^{-1}_{Pl} \,,
\end{eqnarray}
where $H$ is the Hubble parameter. In order to obtain contraction in the string frame
and expansion in the Einstein frame we require $\Theta_{m} < 0$ and $\Theta_{Pl} > 0$. The quantities on the right
hand side of the above eqs. can be written either in the Einstein or Jordan frames as
long as both terms are in the same frame. 

An Anamorphic phase can be obtained in the context of a generalized dilaton gravity action 
(dilaton gravity can be viewed as the low energy limit of string theory - if the antisymmetric 
tensor field of string theory is set to zero).  The action proposed in \cite{an} consists of
a single scalar field $\phi$ non-minimally coupled to the Ricci scalar and non-linearly coupled 
to its kinetic energy density. In addition, we assume the existence of a potential $V_J$
for $\phi$. Specifically, the scalar-tensor theory action of \cite{an} is
\begin{equation} \label{action}
\small S \, = \, \int d^{4}x \sqrt{-g} \left(\frac{1}{2}M_{Pl}^{2}(\phi)R - \frac{1}{2}k(\phi)(\partial_{\mu} \phi)^{2}
- V_{J}(\phi) + \mathcal{L}_{m}\right),
\end{equation}
where $g_{\mu\nu}$ is the metric, $R$ is the Ricci scalar and $k(\phi)$ is the non-linear kinetic coupling 
function. The time dependence of the Planck mass is given by another function $f(\phi)$ via
\begin{equation} 
M_{Pl}(\phi) \, \equiv \, m_{pl}\sqrt{f(\phi)} \, ,
\end{equation}
with $m_{pl}$ being the reduced Planck mass in the frame where $M_{Pl}$ is time independent 
\footnote{Here, we use a different notation than in \cite{an} for the time independent reduced 
Planck mass: $m_{pl}$ in this paper is equivalent to $M_{Pl}^0$ from \cite{an}.}. $V_{J}$ is the potential 
energy density  and $\mathcal{L}_{m}$ is the Lagrangian density of matter and radiation. Unlike what 
was done in \cite{an}, here we will not set the $m_{pl}$ factors to $1$. The action  is written
in the string (or Jordan) frame, and the label J denotes quantities in the Jordan frame. 
 Hereafter we write the Hubble parameter $H$ and perform all our analisys in the Jordan frame. We found
this choice appropiated because in this frame, $\mathcal{L}_m$ and mass terms that might appear in this
contribution are independent of the anamorphic field. Moreover, we are interested in studying the contraction
phase which is better described in Jordan frame.

The non-trivial kinetic coupling and $\phi-$dependence of $M_{Pl}$ in the Lagrangian can 
lead to different signs of $\Theta_{m}$ and $\Theta_{Pl}$ during the Anamorphic phase ($\Theta_{m}<0$ 
and $\Theta_{Pl}>0$). There are other requirements in order to obtain an Anamorphic phase
of contraction: the rate of contracting has to be sufficiently slow to ensure that spatial curvature
and anisotropies are diluted. On the other hand, in the Einstein frame one must obtain an
almost exponential expansion in order to obtain a nearly scale-invariant spectrum of
cosmological perturbations. In addition, one must ensure that the resulting action is
ghost-free (this condition is non-trivial since $k(\phi)$ is negative in the Anamorphic phase).
As shown in \cite{an} these condition are satisfied if 
\begin{eqnarray}
\label{condition}
0 \, < \, 3 + 2k(\phi)\frac{f(\phi)}{(m_{Pl}f_{,\phi})^2} < \epsilon < 1 \,,
\end{eqnarray}
where  $ k(\phi)<0$ during the smoothing phase. The parameter $\epsilon$, is the effective equation of 
state, and is defined as
\begin{equation}
\epsilon \, \equiv \, -\frac{1}{2}\frac{d \, ln \, \Theta_{Pl}^{2}}{d \, ln \, \alpha_{Pl}},
\end{equation}
where $\alpha_{Pl}  \equiv aM_{Pl}/m_{pl}$, with $a$ being the cosmological scale factor. 
Another important auxiliary quantity is $K(\phi)$:
\begin{eqnarray}
K(\phi) \, = \, \frac{3}{2}\left(\frac{f_{,\phi}}{f}\right)^2+\frac{k(\phi)}{f(\phi)m_{pl}^2}\,.
\end{eqnarray}
This quantity must always be positive in the Anamorphic phase \cite{an}.

A complete Anamorphic scenario should describe the Anamorphic contracting phase followed by a
cosmological bounce leading to the hot expanding phase of Standard Big Bang cosmology
during which the Jordan and Einstein frames coincide. It is possible to obtain such a
cosmology using the action (\ref{action}), as described in \cite{an}. In order to
obtain such a boucing cosmology it is important that $k(\phi)$ changes sign.
Since we are interested in obtaining a {\it lower bound} on the total number of particles produced 
during the entire cosmological evolution to the present time, we can consider particle production 
only in the Anamorphic contracting phase. Particle production is due to the squeezing of
the fluctuation modes. Such squeezing happens during the Anamorphic phase and continues
through the bounce phase. If we neglect the squeezing after the end of the Anamorphic
phase we obtain a lower bound on the number of particles produced.

We are going to focus on a specific class of the simple Anamorphic models introduced in \cite{an}, 
where the gravitational coupling during the Anamorphic phase of contraction is 
\begin{equation}
f(\phi) \, = \, \xi e^{2A\phi} \, ,
\end{equation}
and the kinetic coupling is given by 
\begin{equation}
k(\phi) \, = \, -e^{2A\phi} \, .
\end{equation}
Furthermore, the potential is taken to be 
\begin{equation}
V_{J} \, = \, V_{0}e^{B\phi} \, . 
\end{equation}
The parameters $\xi$, $A$, $B$ and $V_{0}$ are positive real numbers. 
For this special case, the equation of state is approximately constant during almost all the 
Anamorphic phase and
\begin{eqnarray}
\epsilon \, = \, \frac{1}{2 K}(B-4A)^2 \,, \qquad K \, = \, 6A^2-\frac{1}{\xi \, m_{pl}^2} \,,
\end{eqnarray}
and hence the condition (\ref{condition}) becomes
\begin{equation}
0 \, < \, \frac{1}{2}(4A-B)^{2} \, < \, K\, < \, A|4A-B| \, ,
\end{equation}
a strict condition on the model parameters, which however can be satisfied.

In the following, we solve for the background cosmology resulting in this class of models.
The Friedmann equation and the equation of motion for $\phi$ yield
\begin{equation} \label{eqSFried}
\Theta_{Pl}(\phi) \, = \, \sqrt{\frac{V_{J}}{m_{pl}^4 f^{2}(3-\epsilon)}} \,, \qquad 
\dot{\phi}_J(\phi) \, = \, \Theta_{Pl} m_{pl}\sqrt{\frac{2\epsilon}{K} f} \,.
\end{equation}
From the above equations, and making use of the definition of $f$, we find that the 
Jordan frame Hubble parameter during the Anamorphic phase is given by:
\begin{align}
 H \, = \, \left(\sqrt{\frac{K}{2\epsilon}}-A\right) \dot{\phi} \, \equiv \, -\alpha_1 \dot{\phi}\,,
 \label{alpha1_def}
\end{align}
where $\alpha_1$ is positive because of condition (\ref{condition}). The Jordan frame Einstein 
equations for a spatially flat universe without anisotropies then take the following
form in the contracting Anamorphic phase  \cite{an}
 \begin{align}
 & 3H^2M_{pl}^2 = \frac{1}{2} k(\phi) \dot{\phi}^2+V_J - 3H m_{pl}^2 \dot{f} \,, \nonumber \\
 & \dot{H} = -\frac{1}{2m_{pl}^2} \frac{k(\phi)}{f} \dot{\phi}^2 +\frac{1}{2} \frac{f_{,\phi}}{f} H\dot{\phi} - \frac{1}{2} \left( \frac{f_{,\phi \phi}}{f} \dot{\phi}^2 
 + \frac{f_{,\phi}}{f} \ddot{\phi} \right) \,.
 \end{align}
 From the above equations it is possible to show that the Hubble parameter in the
 anamorphic phase $t < 0$ evolves as 
 \begin{equation} \label{Hcont}
 H \, = \, p/t, 
 \end{equation}
 with
\begin{equation}\label{pp}
p \, = \, -\frac{\sqrt{\frac{K}{2\epsilon}}-A}{A-\frac{|B-4A|}{2}} \,.
\end{equation}
To obtain Ekpyrotic-type contraction in the Jordan frame we require $0 < p <  1/3$ \footnote{Note
that we do not require $p \ll 1$ but only $\epsilon \ll 1$.}.
In this case, the Anamorphic field comes to dominate the cosmological dynamics
during contraction for a wide range of initial conditions, showing that there is no
initial condition problem. As explained  in \cite{an} this model describes a contracting phase 
that homogenizes, isotropizes and  flattens the universe without introducing initial conditions or multiverse problems and 
unlike in inflation, initial conditions do not have to be finely-tuned. The reason behind this is because the time-varying masses 
suppress the anisotropy.

In Section IV of this paper we will study the energy density generated by Parker
particle production. Since we will be mainly interested in the relative contribution
of these particles to the total energy density, we need to find the expression for
the background energy density as a function of time $t$. Considering that during 
the Anamorphic phase the Anamorphic field is dominating, the background energy 
density is given as a function of  $\Theta_{Pl}$ by
\begin{equation}
\rho_A \, = \, 3\Theta_{Pl}^2 M_{Pl}^4 \, , 
\end{equation}
and,  by  using  Eqs. (\ref{eqSFried}), (\ref{alpha1_def}), (\ref{pp}) and (\ref{Hcont}) we have
\begin{equation}
\rho_A \, \approx \, 3(p - 1)^2 \, f \, \frac{m_{pl}^2}{t^2}\,,
\label{rho_bg}
\end{equation}
where we have used the fact that $\epsilon \ll 1$. From the second Einstein  equation, 
we obtain the following expression for the time evolution of the Anamorphic field:
\begin{equation}
\ddot{\phi} = \left( A - \sqrt{\frac{K(\phi) \epsilon}{2}} \right) \dot{\phi}^2\,,
\label{EOM_phi}
\end{equation}
and given that $\dot{\phi}=-p/t \alpha_1$ we can obtain $\phi(t)$. Thus, considering the 
solution for $\phi(t)$ and the background energy density (\ref{rho_bg}) 
at the end of the  Anamorphic phase, namely at $t = t_{end}$, we have
\begin{equation}\label{thebg}
\rho_{bg}(t_{end}) \, \approx \, 3(p - 1)^2 \, f_{end}\frac{m_{pl}^2}{t_{end}^2}\,,
\end{equation}
where $f_{end}$  is given by $ f_{end}=   \xi \, \mathrm{e}^{2A \phi_(t_{end})}$. 
If we assume that the bounce occurs at the maximum density $\rho_{max} \, = \, M^4$, 
where $M$ is the mass scale of new physics, 
(which we suppose to be between the scale of particle physics "Grand Unification" 
(GUT) and the Planck scale), we can then use $\rho_{bg}(t_{end}) \sim \rho_{max}$ 
and solve (\ref{thebg}) for the time $t_{end}$. 

\section{Cosmological perturbations in the Anamorphic phase}

Since in the Einstein frame the Anamorphic phase is one of almost exponential
expansion, it allows for the generation of a nearly scale-invariant spectrum of 
adiabatic curvature modes and gravitational waves with small non-Gaussianities.  
In fact, the tilts of both the scalar and tensor perturbation spectra are red, whereas
in the Ekpyrotic scenario one obtains a blue nearly vacuum spectrum of the
fluctuations. In the class of examples considered in the previous section the spectral index 
$n_s$ of cosmological fluctuations is given by \cite{an}
\begin{equation}
n_s-1 \, = \, -\frac{(B-4A)^2}{K} \,  = \, -2\epsilon \, ,
\end{equation}
which implies that 
\begin{equation}
n_s \, \approx \, 1
\end{equation}
if $\epsilon \ll 1$. In the following we will compute the power spectrum
of cosmological perturbations at the end of the anamorphic phase,
assuming that the inhomogeneities originate as vacuum fluctuations
on sub-Hubble scales in the far past. 

In the absence of anisotropic stress (which indeed is not present in the
scalar-tensor model which we are considering), it is possible to choose
a gauge in which the perturbed Jordan frame metric takes the form
\begin{equation} \label{ansatz}
ds^{2} \, = \, -dt^{2} + a^2 e^{2\zeta(t,x^{i})} dx^{i}dx_{i} \, ,
\end{equation}
where $\zeta$ is the curvature fluctuation variable in comoving
gauge.

%%RB Check that the above two changes are correct: interpretation
%% of $\zeta$ and of the metric.

Following the usual theory of cosmological perturbation, the equations of
motion for the linear cosmological fluctuations can be obtained by
inserting the above ansatz (\ref{ansatz}) into the full action
and expanding to quadratic order in $\zeta$. The result for the
quadratic terms is \cite{Mukh2, MFB}
\begin{align}\label{acao}
S^{(2)} &=\frac{m_{pl}^2}{2}\int d\eta d^3 x \, 
\alpha_{pl}^2 \epsilon \, [\zeta'^2 -c_{s}^2(\partial_{i} \zeta)^2] \nonumber \\
&=\int d\eta d^3 x \, \frac{z^2}{2} \, [\zeta'^2 -c_{s}^2(\partial_{i} \zeta)^2],
\end{align}
where for scalar field matter the speed of sound is $c_{s}^{2}=1$. Here, $\eta$
is conformal time given by $d\eta = a^{-1}dt$ 
and the prime indicates a derivative with respect to conformal time. 
The only difference compared to the equation for fluctuations in Einstein
gravity with canonically coupled scalar field matter is in the form of the
function $z(\eta)$ which in our model is given by
\begin{equation}
z^2 = 2 \alpha_{pl}^2 \,\epsilon \,m_{pl}^2 = 2 a^2 f \,  \epsilon \,  m_{pl}^2\, ,
\end{equation}
and in the fact that in (\ref{acao}) the scale factor in the integrand is not
$a(t)$, but the Einstein frame scale factor $\alpha_{pl}$ given by
\begin{equation}
\alpha_{pl}(t) \, = \, a(t) \frac{M_{pl}(t)}{m_{pl}} \, .
\end{equation}
In terms of the Mukhanov-Sasaki \cite{Mukh2} variable $v_k =z \zeta_k$
the action is that of a canonically normalized scalar field with time-dependent
mass. The resulting equation of motion is
\begin{equation}\label{eq}
v_{k}'' +(k^2 - \frac{z''}{z})v_{k} \, = \, 0 \,,
\end{equation}
where the comoving momentum is denoted by $k$. 

We must solve this equation in the Anamorphic phase, in which
\begin{equation}
\frac{z''}{z} \, = \, \left[\frac{2+\epsilon}{(1-\epsilon)^2}\right]\frac{1}{\eta^2} \,,
\label{squeezing_parameter}
\end{equation}
where $\epsilon$ is the effective equation of state described in the previous section. 
From this equation, together with the expression for the 
spectral index in our model ($n_{S}-1 \approx -2\epsilon$), we can see that our effective 
equation of state parameter, $\epsilon$, plays an analogous 
role to the slow-roll parameter in inflationary models in that it will determine
the slope of the spectrum \footnote{Experts on cosmological perturbation theory
will immediately see from the factor $2$ in (\ref{squeezing_parameter}) that for $\epsilon \ll 1$
an almost scale-invariant spectrum of fluctuations will result.}.

We now review the computation of the power spectrum.
We will solve equation (\ref{eq}) in a more general setup which can be applied 
to several other scenarios (application to the Ekpyrotic and New Ekpyrotic models was
done in \cite{us}). In this general setup we re-write $z''/z$ as
\begin{eqnarray}\label{nucurv1}
\frac{z''}{z} \, = \, \frac{\nu^2-1/4}{\eta^2} \, .
\end{eqnarray}
For the Anamorphic phase we have (comparing with (\ref{squeezing_parameter}))
\begin{eqnarray}\label{nu}
\nu \, = \, \sqrt{\frac{2+\epsilon}{(1-\epsilon)^2}+\frac{1}{4}} \, .
\end{eqnarray}
In an Anamorphic phase, modes start out with wavelength smaller than the Hubble
length, they cross the Hubble radius $l_H(t) \, \equiv \, H(t)^{-1}$ at a time $\eta_H(k)$, 
and then propagate on super-Hubble scales. The mode evolution on
sub-Hubble and super-Hubble scales is very different. On sub-Hubble scales they
oscillate, while they are squeezed on super-Hubble scales \footnote{The solutions
valid for all times are given by Bessel functions, but introducing these tends to
obscure the physics.}. 

Given a mode with comoving wave number $k$, the conformal time $\eta_H(k)$ associated
with Hubble radius crossing is given by
\begin{eqnarray}\label{hcross}
k^2\eta^2_H(k) \, = \, \nu^2-\frac{1}{4} \,.
\end{eqnarray}
For sub-Hubble modes the  $k^2$ term dominates over the $z''/z$ term and the solution 
to the equation of motion for the perturbations are oscillatory (fixed amplitude). Assuming 
that we start in the Bunch-Davies vacuum, the sub-Hubble solution is
\begin{eqnarray} \label{vacuum}
v_k \, =   \, \frac{e^{-ik\eta}}{\sqrt{2k}} \,.
\end{eqnarray}
On the other hand, for super-Hubble modes the $z''/z$  term dominates over the  $k^2$ term and the perturbations suffer squeezing, resulting in the solution
\begin{eqnarray} \label{vsol}
v_k \, = \, c_1(k) \frac{\eta^{1/2-\nu}}{\eta_H(k)^{1/2-\nu}} 
+ c_2(k) \frac{\eta^{1/2+\nu}}{\eta_H(k)^{1/2+\nu}} \, .
\end{eqnarray}
The coefficients $c_1(k)$ and $c_2(k)$ of the two modes can be found by matching 
$v_k$ and $v'_k$ at Hubble radius crossing $\eta_H(k)$. This yields
\begin{eqnarray}
c_1(k) \, &=& \, \frac{1}{2\nu}\frac{1}{\sqrt{2k}} e^{- i k \eta_H(k)} \left[ \nu +\frac{1}{2} + i k \eta_H(k) \right] \, , \nonumber \\ 
c_2(k)  \, &=& \, \frac{1}{2\nu}\frac{1}{\sqrt{2k}} e^{- i k \eta_H(k)} \left[ \nu -\frac{1}{2}- i k \eta_H(k) \right] \, .
\label{coeffs}
\end{eqnarray}
For the value of $\nu$ corresponding to the Anamorphic phase, the first mode is growing,
and the second decaying.

The solution for $v_{k}$, given by the above equations, can be used to compute the power 
spectrum predicted by the model, as we will study in the following subsection. It can
also be used to estimate Parker particle production in the Anamorphic phase, as we discuss
in Section IV.

\subsection{Power Spectrum}

In this subsection we evaluate the power spectrum of the curvature perturbations produced 
during the Anamorphic phase. We are interested in modes which are super-Hubble
at the end of the Anamorphic phase. In this case we can neglect the contribution of
the decaying mode in (\ref{vsol}) and we obtain 
\begin{equation}
|v_k(\eta_{end})|^{2} \, = \, \frac{1}{4\nu^{2}} \frac{1}{2k}[(\nu+ 1/2)^{2} + k^{2}{\eta_{H}}^{2}] \left|\frac{{\eta_{end}}^{1/2-\nu}}{{\eta_{H}}^{1/2-\nu}}\right|^2\,.
\end{equation}
Therefore,  since $\nu\approx3/2$ (which can be obtained from eqs. \ref{squeezing_parameter} and \ref{nucurv1} considering $\epsilon \approx 0$) and by using (\ref{hcross}) this reduces to
\begin{equation}
|v_k(\eta_{end})|^{2} \, = \, \frac{2}{3k^{3}\eta_{end}^{2}} =  \frac{(1-p)^2}{p^2} \frac{2H^2_{end}}{3k^{3}}  \,.
\end{equation}

With this result we can evaluate the dimensionless power spectrum of curvature fluctuations
which for $\nu = 3/2$ (i.e. $\epsilon = 0$) is given by 
\begin{equation}
\mathcal{P}_{\zeta} \, = \, \frac{k^{3}}{2}|\zeta_{k}(\eta)|_{end}^{2} \, = \, \frac{k^{3}}{2} |v_{k}^{2}/z^2 |_{end} \, ,
\end{equation}
where $\zeta_k = v_k/z$, and $z^2_{end}=2 f_{end} \, \epsilon \, m_{pl}^2$. 
Therefore, we obtain the following power spectrum
\begin{equation}
P_{\zeta} \, = \, \frac{\left( 1/p -1 \right)^2}{3 f_{end} } \, \frac{H^2_{end}}{2\, \epsilon\,  m^2_{pl}} 
\, = \, \beta^2 \, \frac{H^2_{end}}{2\, \epsilon\,  m^2_{pl}}\,.
\end{equation}
In this limit we obtain an exactly scale invariant spectrum. For $0 < \epsilon \ll 1$
the spectrum obtains a slight red tilt, in the same way that a slight red tilt emerges
for simple slow-roll inflation models (the parameter $\epsilon$ in the Anamorphic 
phase of this model plays the same role as the slow-roll parameter in the inflationary scenario). 

The amplitude of the power spectrum of cosmological perturbations is given by
a very similar expression as in inflationary cosmology in terms of the dependence
on the Hubble parameter and on $\epsilon$. There is a difference in the overall amplitude
which is given by the multiplicative factor $\beta$ which depends on two parameters of the 
model, $p$ and $f$ evaluated at the end of the Anamorphic phase. The presence of
this non-trivial factor in the amplitude will imply that the relative importance of Parker
particle production of matter particles will be different in the Anamorphic scenario
than in inflationary cosmology.

%%RB: Is this correct?

To fix the model parameters we  compare the power spectrum at the end of the 
Anamorphic phase with the observed value as measured by observations of the 
Cosmic Microwave Background (CMB). The anamorphic contraction is 
followed by a second contracting phase in which $\epsilon$ varies considerably, and
then by the bounce phase (the reader is refereed to fig.
2 in \cite{an} for an sketch the overall behaviour of the parameters during the contracting
phases). Unlike in \cite{an} in which the so called anamorphic phase
last until $\epsilon \approx 1$, here we define the anamorphic phase as the phase in which $\epsilon \approx const$
. In principle, the power spectrum could also grow
during these two phases. We are here neglecting any
such additional growth of fluctuations. If both the pre-bounce phase when
$\epsilon$ undergoes rapid change and the bounce phase
are short, then this will be an excellent approximation, as shown in detailed
studies of the evolution of fluctuations in other nonsingular bounce models
\cite{other}. If the two additional phases are long (on a time scale set by
the maximal value of $|H|$), then the conditions on the model parameters
which we derive below in order to obtain agreement with the observed power
spectrum will change. What will, however, not change is the relative amplitude
of Parker particle production and cosmological perturbations since both will be
effected in the same way in the two additional phases. 

According to the CMB observations (the latest results being from the \textit{Planck} team 
\cite{planck}), the amplitude of the power spectrum is $A_{\zeta}\approx 10^{-10}$.  
Written in terms of the above power spectrum this yields
\begin{equation}\label{ampl}
A_{\zeta} \, \sim \, 10^{-10} \, \sim \, \frac{\beta^2}{3} \frac{1}{2 \epsilon\,} \frac{M^4}{m^4_{pl}}\,,
\end{equation}
where we used the fact that $3 H^2_{end}/m^2_{pl}= \rho_{bg}/m^4_{pl}$ and that the 
background energy density is equal to the maximum density, $\rho_{max} = M^4$, 
where $M$ is the mass scale of new physics. Given a value of $M$ we can then constrain the 
parameter $\beta$, like was done for the new Ekpyrotic model in \cite{us}. Given that we 
know that $0 < p < 1$, this translates into a constraint on the parameter 
$f_{end}$ that enters in the expression for $\beta$:
\begin{equation}
f_{end} \sim \frac{1}{10^{-10}} \left(\frac{1}{p} -1 \right)^2 \frac{1}{18 \epsilon} \, \frac{M^4}{m^4_{pl}}\,.
\label{fend}
\end{equation}
Since the value of $\epsilon$ must have the same value as in the inflationary scenario 
$\sim 10^{-2}$ to yield agreement with the observed slope of the spectrum of
cosmological perturbations, we obtain bounds on $f_{end}$ which depend on the ratio of the 
New Physics and Planck scales.

\section{Parker particle production in Anamorphic phase}

In this section we are interested in computing the number and energy densities of
particles created by squeezing of the modes of a matter scalar field $\chi$ (which we
treat as a test scalar field minimally coupled in the Jordan frame)
by the time of the end of the Anamorphic phase.

Particle production is a consequence of squeezing. Hence, there are no
particles generated for modes with wavelengths smaller than the Hubble length
at the end of the anamorphic phase. For super-Hubble modes we must be
careful how to interpret the squeezed vacuum state in terms of particles.
The particle interpretation only becomes valid when the modes re-renter
the Hubble radius during the Standard Model phase of expansion. As we
will see, the energy density in produced particles falls off in the infrared.
The integral over modes is hence dominated by modes which have a
wavelength only slightly larger than the Hubble radius at the end of the
anamorphic phase. These are modes which re-enter the Hubble radius
shortly after the beginning of the Standard Model phase of expansion.
Hence, the particles energy density which we here compute indeed has
an interpretation as particle energy density beginning very early in the
expanding phase.

As mentioned at the end of the previous section, what we are computing
here is a lower bound on the particle energy density since we are
neglecting the squeezing during the bounce and in the phase when
$\epsilon$ is rapidly changing before the bounce. However, if the
squeezing in those phases were important, it would also be important
for the cosmological fluctuations, and the squeezing at the end
of the anamorphic phase would have to be smaller. The bottom line
is that the effects of the decrease in squeezing at the end of the
Anamorphic phase and the extra squeezing after the end of the
Anamorphic phase will counteract, and will not substantially effect
our final result. Just like in \cite{an} we considered that the power spectrum of cosmological perturbations
must be computed in the anamorphic phase. However if that
was not the case, then the amplitude of cosmological perturbations would increase in
the second phase. This only means that we would have made an unnapropriated choice
of the normalization of the spectrum. By fixing the normalization to the correct one
(corresponding to a smaller amplitude) and considering the increase in the following
phase to the observed value would imply in the same final result as we obtained. Since
the Parker particle production and the cosmological perturbations are affected in the
same way in the two additional phases the same conclusion is valid for the density of
the particles produced.

The squeezing of the mode functions in a time-varying cosmological background 
corresponds to gravitational particle production \cite{Parker} (see e.g.
\cite{Birrell:1982ix, Mukhanov:2007zz} for textbook treatments). As is
the standard approach in quantum field theory on curved space-times,
the modes $\chi_k$ of a test scalar field $\chi$ (which can also be the
cosmological fluctuations or gravitational waves themselves) can be 
expanded into positive and negative frequency modes. As described in \cite{Parker}, 
initial pure positive or negative frequency modes (we are using the Heisenberg 
representation) whose coefficients are interpreted as creating and annihilation
operators become mixed during the time evolution. This correspondes
to particle production. 

At any time $\eta$, the mode functions of a full solution of the equations of motion 
can be expanded momentarily \footnote{This means that the field values and their
first time derivatives coincide.} into a linear combination of 
the instantaneous vacuum solutions $\chi_{v,k}$, i.e. in terms of a local positive and 
negative frequency modes.
\begin{align}
\chi_k (\eta) = \alpha_k \chi_{v,k} (\eta) + \beta_k \chi_{v,k}^* (\eta) \nonumber \,, \\
\chi'_k(\eta) = \alpha_k \chi'_{v,k}(\eta) + \beta_k \chi'^*_{v,k} (\eta)\,.
\label{bogoliubov_modes}
\end{align}
%\begin{align}
%v_k (\eta) = \alpha_k v_{BD} (\eta) + \beta_k v_{BD}^* (\eta) \nonumber \,, \\
%v'_k(\eta) = \alpha_k v'_{BD}(\eta) + \beta_k v_{BD}'^* (\eta)\,.
%\label{bogoliubov_modes}
%\end{align}
The time-independent coefficients are the Bogoliubov coefficients that, for bosons satisfy 
$|\alpha|^2_k - |\beta|^2_k = 1 $, if both sets of modes are normalized. The quantity 

\begin{equation}
n_k \, \equiv \, |\beta_k|^2 
\end{equation}
is interpreted by a late adiabatic time observer as the particle number in the mode $k$ which 
has been produced starting from a vacuum initial state. The squeezing which $v_k(\eta)$ 
undergoes leads to a growth of the expansion coefficients and hence to a growth in the
number density $n_k$.  

In the following, we shall consider two types of particles. First we will consider 
particles associated to the adiabatic fluctuation mode (this will be particles
associated with the Anamorphic scalar field). Then, we will consider $\chi$ as a 
massless matter field minimally coupled in the Jordan frame. This corresponds
to usual matter.

\subsection{Adiabatic Mode Particles}

In this subsection, we are going to study the energy density in particles
associated with the adiabatic fluctuation mode, i.e. $\phi$ particles.
We use the solution for the mode function $v_k$ from the previous 
section, i.e. $\chi_k \equiv v_k$, given by (\ref{vsol}) and (\ref{coeffs}), that represents the 
modes that suffered squeezing after crossing the Hubble radius. Using (\ref{bogoliubov_modes}), 
we obtain the Bogoliubov coefficients by considering that the solution and its derivative can be 
expanded in terms of the Bunch-Davis vacuum basis (Eq. (\ref{vacuum})).

Considering only the growing solution of (\ref{vsol}) on super-Hubble scales, we can evaluate 
the Bogoliubov coefficient $\beta_k$ and obtain
\begin{eqnarray} \label{betaeq}
\beta_{k}(\eta) \, = \, \frac{c_1(k) \sqrt{2k}}{2} \left( \frac{\eta}{\eta_H(k)} \right)^{1/2 - \nu} 
\left[ 1 + \frac{1/2 - \nu}{i k \eta} \right] \, .
\end{eqnarray}
Recalling the expression  for $\nu$ (Eq.(\ref{nu})) in the limit $\epsilon \ll 1$, i.e $\nu \approx 3/2$,
and substituting the coefficient from (\ref{coeffs}), we have that $\beta_{k}$ is given by
\begin{eqnarray}
\beta_k(\eta) \,&=& 
\frac{1}{3}e^{- i k \eta_H(k)} \left[1 + i \frac{k \eta_H(k)}{2}\right] \left( \frac{\eta}{\eta_H(k)} \right)^{-1} \, . \nonumber
\end{eqnarray}
The number density of produced particles is hence given by
\begin{eqnarray}
n_k(\eta) \, = \, \frac{1}{9}\left[1 + \frac{(k \eta_H(k))^2}{4}\right] \left( \frac{\eta}{\eta_H(k)} \right)^{-2} \, .
\end{eqnarray}
Note that this result is the same as we obtained in the New Ekpyrotic scenario \cite{us}. 

We can  now compute the energy density of the particles produced \footnote{We recall
that the particle interpretation is only valid after the time of Hubble radius re-entry.} 
until the end of the Anamorphic phase, i.e. the conformal time $\eta_{end}$ given by 
$\rho_{p}(\eta) \sim \left( (1-p)^4 /p^2 \right) H^2_{end} /\eta_{end}^2$
(here we have normalized the scale factor to give $a_{end}$=1). Thus, at $t=t_{end}$
\begin{eqnarray}
\rho_p(t_{end}) \, \sim \, (1-p)^4 \, t^{-4}_{end}\,.
\end{eqnarray}
%%

%%RB: I don't understand where the coefficient on the rhs comes from.

We can compare this density of produced particles with the background density 
(see eq. (\ref{thebg})) in order to see if the particles created are sufficient to lead to a 
post bounce hot big bang phase
\begin{equation}
\frac{\rho_p(t_{end})}{\rho_{bg}(t_{end})} \, \sim \, \left( \frac{1}{p} -1 \right)^2 \frac{H^2_{end}}{3 f_{end} \, m^2_{pl}} 
\, = \, \left( \frac{1}{p} -1 \right)^2 \frac{1}{9\, f_{end}} \frac{M^4}{m^4_{pl}}\,.
\label{ratio_density}
\end{equation}
Given the constraint on $f_{end}$ in order to respect the CMB constraint on the amplitude 
of the power spectrum we can see that, substituting Eq. (\ref{fend}) 
in the above ratio we get
\begin{equation}
\frac{\rho_p(t_{end})}{\rho_{bg}(t_{end})} \, \sim \, \frac{10^{-12}}{\epsilon} \, \sim \, 10^{-10}\,.
\end{equation}
We can see that this ratio is smaller than the one one would get from inflation, since in inflation this is proportional to $M^4_{GUT}/m^4_{pl} \sim 10^{-12}$. Thus,
as in the case of inflation, in this model the energy density in particles created by 
Parker particle production of the Anamorphic field $\phi$ during the Anamorphic phase is not sufficient
to reheat the universe.

\subsection{Matter Field Dynamics}

In the last section we computed the density of particles produced from the curvature 
perturbations whose main contribution comes from the dominant Anamorphic field. 
Now let us compute the density of particles produced via the Parker mechanism for 
a massless matter field minimally coupled to gravity in the Jordan frame in a background 
driven by the Anamorphic field. This field stands for the matter of the Standard Model
of particle physics. It is hence the field of most interest regarding reheating  of the universe.

In the Jordan frame, the mass of the matter component is independent of the Anamorphic field 
and is constant \cite{an}. The dynamics of this field is different from that of
the gravitational perturbations since the squeezing term in the equations for the Fourier modes of the matter field 
and of the perturbations, associated to the Mukhanov-Sasaki variable, are 
different. After rescaling by the Jordan frame scale factor, the 
Fourier modes of the matter field obey the  equation
\begin{equation}\label{matter}
\chi''_{k} + \left(k^2 -\frac{a''}{a}\right)\chi_{k} \, = \, 0 \,,
\end{equation}
where during the Anamorphic phase
\begin{equation}
\frac{a''}{a} \, = \, -\frac{p(1-2p)}{(1-p)^2}\frac{1}{\eta^2}\,.
\end{equation}
Analogous to eq. (\ref{nucurv1}), we can write the time dependent part of the effective mass
as
\begin{equation}
\frac{a''}{a} \, \equiv \, \frac{\nu_m^{2}-1/4}{\eta^{2}}\,,
\label{nu_m}
\end{equation}
which implies that
\begin{equation}\label{nup}
\nu_m \, = \, \sqrt{\frac{1}{4}-\frac{p(1-2p)}{(1-p)^2}} 
\end{equation}
Note that for the range $0<p<1/3$ of values of $p$ which give Ekpyrotic
contraction $\nu_m$ is always real.  With that, the equation of motion takes 
the same form as discussed in the previous sections
\begin{equation} \label{final}
\chi''_{k} + \left[k^2 -\frac{\nu_m^2-1/4}{\eta^2}\right]\chi_{k} \, = \, 0\,.
\end{equation}

We see that for the entire range of values of $p$ of interest, there is less
squeezing for these mode functions than for the cosmological perturbations
since the coefficient of the $\eta^{-2}$ term in (\ref{final}) ranges from $0$ to
$1/4$ whereas it is $2$ in the case of cosmological perturbations. This
implies that if we start with vacuum fluctuations, the resulting spectrum
of matter perturbations will be blue. However, since for particle production
the dominant contribution comes from modes which exit the Hubble radius
just before the end of the Anamorphic phase, the fact that we have a blue
spectrum will not in itself lead to a suppression of the energy density in
matter particles relative to the energy density in Anamophic particles
computed in the previous subsection. What will lead to a suppression
of matter particle energy density relative to Anamorphic particle energy
density is the fact that the effective Hubble radius crossing conditions
differ. For matter particles, it follows from (\ref{final}) that the Hubble
crossing condition is
\begin{equation}
k^2 \eta_H(k)^2 \, = \, \nu_m^2 - \frac{1}{4} \, \sim \, p ,
\end{equation}
instead of
\begin{equation}
k^2 \eta_H(k)^2 \, \sim \, 2 ,
\end{equation}
as is the case for the cosmological perturbations \footnote{Hopefully the
reader will forgive us for using the same symbol $\eta_H(k)$ for the
two different calculations.}. Hence, the value of $k_H$ is suppressed
by $\sqrt{p}$ for matter fluctuations. Since the energy density
is dominated by this ultraviolet scale, it leads to a suppression of
the matter energy density relative to the Anamophic particle energy
density.

The analysis of particle production parallels the discussion in the
previous subsection. We begin on sub-Hubble scales with the 
Bunch-Davies vacuum solution, 
\begin{equation}
\chi_k \, = \, e^{-ik\eta}/(\sqrt{2k}) \, ,
\end{equation}
and while $k^2 \ll a''/a$ the solution is the same as Eqs. (\ref{vsol}) 
and (\ref{coeffs}). Thus the Bogoliubov coefficients $\beta_k$ and 
the number density of produced particles $n_k$ are the same
that in previous case but with $\nu_m$ instead $\nu$. Then
\begin{equation}
n_k(\eta) \, = \, \frac{1}{4\nu^2_m}\left[\frac{(\nu_m+1/2)^2}{4} + \frac{(k\eta_H)^2}{4}\right]
\left(\frac{\eta}{\eta_H}\right)^{1-2\nu_m}\,,
\end{equation}
which leads to following matter density 
\begin{equation}
\rho_p^m \, = \, \frac{1}{(2\pi)^3}\int^{k_H}_{k_0}n_k\,k d^3k \, .
\end{equation}
This integral is dominated in the ultraviolet, and hence we
can extend the integration to $k = 0$. Thus, at the end of Anamorphic phase:
\begin{equation}
\rho_p^m(\eta_{end}) \, \simeq \, A(\nu_m) \,\eta^{1-2\nu_m}_{end} k^{5-2\nu_m}_H \,,
\end{equation}
with
\begin{equation}
A(\nu_m) = \frac{1}{16\pi^2}\frac{(\nu^2_m-1/4)^{\nu_m-1/2}}{5-2\nu_m}\left[1+\frac{1}{2\nu_m}\right] \, .
\end{equation}
Inserting the Hubble radius crossing condition we obtain the following solution 
for the density of particles produced,
\begin{equation}
\rho_p^m(t_{end}) = A(\nu_m) (1-p)^{2\nu_m-1} p^{5/2-\nu_m}  \,t^{-4}_{end}\,,
\end{equation}
which is suppressed compared to the contribution of anamorphic particles for
the reason discussed in the previous paragraph.

Just like in the previous section, we can now write the ratio between the density of particles 
produced and the background density in the end of the Anamorphic phase, 
which is given by
\begin{equation}
\frac{\rho^m_{p}(t_{end})}{\rho_{bg}(t_{end})} \, \approx \, 
A(\nu_{m})(1-p)^{-3+2\nu} p^{1/2 -\nu} \frac{H^2_{end}}{3 f_{end} m^2_{pl}}  \, 
= \,  A(\nu_{m})(1-p)^{-3+2\nu} p^{1/2 -\nu}\frac{1}{9 f_{end}} \left(\frac{M}{m_{pl}}\right)^4 
\end{equation}
and substituting our constraint (\ref{fend}) we get
\begin{equation} \label{final2}
\frac{\rho^m_{p}(t_{end})}{\rho_{bg}(t_{end})} \, \approx \, 2\times 10^{-12}\tilde{A}(p), \qquad \, \tilde{A}(p) 
\, = \, A(\nu_{m}) \left(\frac{\sqrt{p}}{1-p}\right)^{5-2\nu_{m}} \,.
\end{equation}

After the Anamorphic phase a second phase of contraction takes place in which $\epsilon$ 
increases considerably until $\epsilon=3$. During the beginning of this phase $|H|$ keeps increasing until it reaches its maximum value. After that $|H|$ starts decreasing until it 
reaches $H = 0$ at the bounce (just after the second phase ends). 
During the bounce phase $k(\phi)$ changes from negative to positive and the quantity 
$\Theta_{m}$ also  passes through zero and becomes positive. The $\Theta_{m}$ bounce, 
which does not require a $\Theta_{Pl}$ bounce, can occur without violating 
Null Energy Condition (NEC). After the bounce, the field $\phi$ settles at a 
minimum of the potential, $f$ and $k$ become fixed and we have $M_{Pl}\Theta_{m}=M_{Pl}\Theta_{Pl}=H_{E}$, 
in agreement with  Standard Big Bang evolution in Einstein gravity. 

One could argue that during the pre-bounce and bounce phases following 
after the Anamorphic contracting phases the ratio $\rho_{pm}/\rho_{bg}$ 
could increase. However, as already mentioned earlier, if there were enhancement 
of the density produced by Parker mechanism in this 
second phase, the amplitude of the power spectrum would also increase in this period, 
and the value calculated here and specially constraint (\ref{fend}) would have to be
adjusted. These are counteracting effects and will tend to balance themselves
out. Also, if the additional phases are short compared to the $|H|_{max}^{-1}$, the
the additional squeezing will have a small effect. Hence, we consider our
estimate  (\ref{final2}) to give a reliable guide.

The bottom line is that matter particle production via the Parker mechanism is,
as in the case of inflationary cosmology, not able to effectively reheat the
universe quickly. If no additional mechanism is introduced to drain the energy
from the Anamorphic field condensate, then the Universe at the beginning of
the expanding phase will be dominated for a long time by this condensate.
Mechanisms to drain the energy in the condensate include nonlinearities in
the action for $\phi$ or couplings between $\phi$ and regular matter.

\section{Conclusions} 

The  Anamorphic cosmology  recently proposed in \cite{an} corresponds to a new scenario for the early universe which has the interesting property of combining elements and advantages of both the inflationary and the Ekpyrotic scenarios.
In the present work we made a further analysis of a realization of this model. 

We have studied the growth of cosmological perturbations 
in the Anamorphic scenario. After reviewing the background 
dynamics for a specific class of scalar-tensor theories introduced in \cite{an}
we have obtained the constraints on the model parameters obtained by demanding
that the amplitude of the spectrum of cosmological perturbations agree with
observations.

We then studied Parker particle production during the Anamorphic phase of
contraction. We studied both the production of Anamorphic particles and
of test matter particles (particles minimally coupled in the Jordan frame).
We found that, as in the case of inflationary cosmology, Parker particle
production is not effective enough to drain a sizeable fraction of the
energy density from the Anamorphic field condensate. Thus, as in the
case of inflationary cosmology, a new mechanism is required if we want
the universe to be dominated by regular matter close to the beginning of
the phase of expansion.

\section*{Acknowledgement}
\noindent
We thank Anna Ijjas and Paul Steinhardt for comments on our draft.
One of us (RB) wishes to thank the Institute for Theoretical Studies of the ETH
Z\"urich for kind hospitality. RB acknowledges financial support from Dr. Max
R\"ossler, the Walter Haefner Foundation and the ETH Zurich Foundation, and
from a Simons Foundation fellowship. The research of RB is also supported in
part by funds from NSERC and the Canada Research Chair program. WSHR and LG are grateful 
for the hospitality of the Physics Department of McGill University. 
WSHR was supported by Brazilian agency CAPES (proccess No 99999.007393/2014-08).
EF acknowledge financial support from CNPq (Science Without Borders). LG acknowledge financial support from CNPq (Science Without Borders) and CAPES (88887.116715/2016-00).

\end{document}